# Artifact Free Transient Near-Field Nanoscopy

A. J. Sternbach,[1,2] J. Hinton,[2] T. Slusar,[3] A. S. McLeod,[1] M. K. Liu,[4] A. Frenzel,[2] M. Wagner,[2] R. Iraheta,[2] F. Keilmann,[5] A. Leitenstorfer,[6] M. Fogler,[2] H.-T. Kim,[3] R. D. Averitt,[2] and D. N. Basov[1,2,*]

[1]Department of Physics, Columbia University, 116th St and Broadway, New York, NY 10027, USA

[2]Department of Physics, University of California San Diego, 9500 Gilman Dr., San Diego, CA 92037, USA

[3]IT Convergence and Components Laboratory, Electronics and Telecommunications Research Institute, Daejeon 305-350, Korea.

[4]Department of Physics, Stony Brook University, Stony Brook, New York 11794, USA

[5]Ludwig-Maximilians-Universität and Center for Nanoscience, 80539 München, Germany

[6]Department of Physics and Center for Applied Photonics, University of Konstanz, D-78457 Konstanz, Germany

*corresponding author: db3056@columbia.edu

**Abstract:** We report on the first implementation of ultrafast near field nanoscopy carried out with the transient pseudoheterodyne detection method (Tr-pHD). This method is well suited for efficient and artifact free pump-probe scattering-type near-field optical microscopy with nanometer scale resolution. The Tr-pHD technique is critically compared to other data acquisition methods and found to offer significant advantages. Experimental evidence for the advantages of Tr-pHD is provided in the Near-IR frequency range. Crucial factors involved in achieving proper performance of the Tr-pHD method with pulsed laser sources are analyzed and detailed in this work. We applied this novel method to time-resolved and spatially resolved studies of the photo-induced effects in the insulator-to-metal transition system vanadium dioxide with nanometer scale resolution.

## 1. Introduction

Ultrafast optical techniques provide access to processes that occur with awesome rapidity, enabling novel routes to control and interrogate the complex energy landscapes of materials at the focus of modern condensed matter physics [1,2]. Ultrafast techniques have provided insights into coherent motions at atomic length scales [3], excitation or interrogation of selective electronic, lattice, spin or magnetic modes [4,5,6], and domain growth [7]. In materials where multiple degrees of freedom compete ultrafast studies have allowed researchers to identify the degrees of freedom associated with emergent phenomena [8,9,10]. Additionally, ultra-short light pulses have granted access to hidden states of matter [11, 12], creating novel opportunities for material discovery and control.

In the case of quantum materials with strong electronic correlations spatial complexity across phase transition boundaries demands that measurements be performed with nanometric spatial resolution [13]. This approach is needed to map phase inhomogeneities which are thought to play a fundamental role in emergent behavior of a broad class of quantum materials including, but not limited to: colossal magneto-resistance manganites [14], Cu and Fe-based High-Tc superconductors [15,16] and transition metal oxides [17-19]. Merging ultra-fast techniques with ultra-high spatial resolution both brings the unique merits of ultrafast measurements to the nanoscale and enables the exploration of connections between spatial and temporal responses at extreme small time and length scales [20-23]. It is therefore imperative to develop advanced tools for time-resolved investigation at the nanoscale.

Scattering type near-field optical microscopy (s-SNOM) is well suited for-spectroscopy and imaging with 10-20 nm spatial resolution. The spatial resolution afforded by this method is independent of the wavelength of radiation used [24]. A number of works, where nano-Fourier Transform Infrared (FTIR) spectroscopy [25-29] and Electro-optic sampling (EoS) [28] were used have provided a robust demonstration of the potential to couple ultrafast lasers to s-SNOMs to successfully circumvent the diffraction limit and gain access to pure time-resolved information at the nanoscale. Recent results have also shown the strong potential to perform rapid time-resolved nano-imaging with s-SNOM [30,31], which enables a detailed exploration of the role of inhomogeneities across quantum phase transitions in complex materials. All these results have demonstrated that ultrafast s-SNOM is a powerful technique with a bright future [25-33].

One potential difficulty in s-SNOM measurements is that a large contribution from background radiation, which stems from light that scatters off of the tip and/or sample, is present. Decades of experience with s-SNOMs gained by the nano-optics community have identified experimental practices that allow one to suppress the contribution of background radiation and thereby acquire genuine near-field data, which are guaranteed to be artifact-free. One potent approach for eliminating the impact of background radiation is the pseudoheterodyne detection (pHD) method [34]. However, the pHD method is yet to be adapted to pulsed laser sources. In this work we provide the first demonstration of utilizing the pHD method with pulsed laser sources. Based on extensive analysis and modeling we conclude that pHD acquisition is imperative in specific cases that are detailed in section 4. We then present the first results for a prototypical insulator-to-metal transition system $VO_2$, gathered with Tr-pHD with a probe wavelength near 1.5 μm and a pump wavelength near 780 nm. These data are free from the ill influence of background radiation and set the stage for future spatio-temporal exploration of quantum materials at the nano-scale.

## 2.  Overview of Time-Resolved Near-Field Techniques

The aim of this work is to develop a proper framework for time-resolved near-field measurements and to critically evaluate various possible data acquisition protocols. Recent results employing approaches based on Scanning Tunneling Microscopy (STM) [22,23] and

Atomic Force Microscopy (AFM) [25-33] have recently been demonstrated. In this work we focus on the all-optical approach of coupling infrared lasers to an AFM. We begin by discussing the experimental components that are needed to gain access to genuine time-resolved near-field information (Fig.1a).

The centerpiece is an AFM with the tip of the cantilever illuminated with infrared lasers. The metallic AFM probe is polarized by incident light, and together with its mirror image in the sample, generates an evanescent electric field that is confined to the radius of curvature of the tip (10-20 nm); a feat that stems from the near field coupling between the tip and the sample. The AFM tip is then re-polarized by the tip-sample interaction and radiation is scattered into the far-field [24,35]. This radiation, which contains background radiation as well as radiation produced by the near-field interaction, is then sent to a detector. The backscattered radiation from the AFM is usually superimposed with light from a reference arm in order to form an interferometric receiver. Interferometry eliminates the multiplicative contribution of diffraction limited background radiation and provides phase information - as will be detailed below [34,36,37]. Since the voltage generated by common detectors, $u$, is proportional to the light intensity rather than its electric field, we consider the square of the sum of all electric fields:

$$u \propto |\tilde{E}_{ref} + \tilde{E}_{BG} + \tilde{E}_{NF}|^2 \quad (1)$$

Where $\tilde{E}_{ref}, \tilde{E}_{BG},$ and $\tilde{E}_{NF}$ are the electric field phasors, respectively, of the reference arm, the background contribution as well as radiation scattered from the near-field. To experimentally eliminate terms, which do not contain near-field information, the well established tapping technique [38] is commonly used. Within this approach, all terms which are not proportional to $\tilde{E}_{NF}$ can be made negligible (Appendix A). Thus, when the tapping technique is used the detected intensity contains only the terms:

$$u_n \propto \tilde{E}_{ref}^* \cdot \tilde{E}_{NF} + \tilde{E}_{BG}^* \cdot \tilde{E}_{NF} + \tilde{E}_{NF}^* \cdot \tilde{E}_{NF} + \text{C.C.} \quad (2)$$

Typically, the amplitude of the electric field phasors from the background and reference arm are orders of magnitude larger than that from the near-field. Thus the last term in Eq. (2), $\tilde{E}_{NF}^* \cdot \tilde{E}_{NF}$, and it's complex conjugate, are negligible and will not be considered in the remainder of this paper. If no reference arm is used, only the terms $s^{sHD} \propto \tilde{E}_{BG}^* \cdot \tilde{E}_{NF} + C.C.$ are measured, which is generally referred to as the self-homodyne detection (sHD) method. The sHD signal is proportional to the background field, which introduces the so called "multiplicative background" contribution (sections 3, 4, Appendix A). If the reference arm is added, we are left with $s^{HD} \propto \tilde{E}_{ref}^* \cdot \tilde{E}_{NF} + \tilde{E}_{BG}^* \cdot \tilde{E}_{NF} + C.C.$, which is often described as the homodyne detection (HD) method. The HD signal is background-free provided the amplitude of the reference field is much stronger than that of the background field, $|\tilde{E}_{ref}| \gg |\tilde{E}_{BG}|$ [39]. If the latter inequality is not fulfilled, the HD signal is influenced by the background contribution. To totally eliminate the multiplicative background contribution the so-called pseudoheterodyne detection method (pHD) (sections 5, 6) has been devised, which leaves only the term $s^{pHD} \propto \tilde{E}_{ref}^* \cdot \tilde{E}_{NF} + C.C.$ [34]. Results generated using pHD are not influenced by background radiation and are, therefore, guaranteed to be background free. By raster scanning the sample while keeping the positions of the AFM and optics fixed, one is able to extract signal from sHD, HD, or pHD on a pixel-to-pixel basis and construct an image.

In order to gain access to time-resolved information we use pulsed laser sources both for the probe and the pump channels shown schematically in Fig. 1a. Radiation from one channel is sent to the AFM to probe the sample's momentary state in the near-field, (purple in Figure 1a). A second illumination channel is used to pump (or perturb) the sample at a well controlled time delay, $\Delta t_{ps}$, preceding the probing event; the role of the pump is to transiently alter the properties of the sample (red in Fig.1a). We utilize two digital boxcars [40] to measure the pump-induced change of the near-field signal (Section 4; Appendix B), by collecting simultaneously the signals $s_R^X$, just before, and $s_P^X$, at $\Delta t_{ps}$ after the pump pulses. We then plot the difference $\Delta s^X = (s_P^X - s_R^X)/s_R^X$ which is non-zero only if the pump transiently modifies the response of a sample at a given pixel. In our notation, the upper-script assigns the method of time resolved detection, i.e. Tr-pHD for pseudo-heterodyne detection, Tr-sHD for self-homodyne detection, and Tr-HD for homodyne detection. Each of these methods has the capacity to produce reliable time-resolved information at the nanoscale under certain conditions. The fidelity of data against various potential artifacts provided by each of these methods will be critically evaluated in this work.

## 3. Experimental Comparison of Data Acquisition Methods in Time-Resolved Near-Field Measurements

In this work we investigated thin films of vanadium dioxide ($VO_2$): a correlated electron material that undergoes an insulator-to-metal (IMT) transition above room temperature. The highly oriented $VO_2$ films on $[001]_R$ $TiO_2$ substrate, as well as polycrystalline samples on $Al_2O_3$ substrates, were fabricated by the pulsed-laser deposition method; details of thin film fabrication and characterization have been reported elsewhere [17, 41]. Static Near-Field imaging works have shown that $VO_2$ experiences a percolative phase transition with co-existing insulating and metallic states in the vicinity of the IMT [17, 19, 41]. The transition temperature of $VO_2$ films can be tuned by epitaxial strain [42]. In general, compressive (tensile) strain along $c_R$ yields $T_{IMT}$ lower (higher) than in bulk [41-43]. Bulk crystals and unstrained polycrystalline films on sapphire substrates usually have an IMT close to $T_{IMT}$ =340 K. Films on $[001]_R$ $TiO_2$ substrate are compressively strained along $c_R$, leading to a $T_{IMT}$ < 340 K [41]. Topographic corrugations, or "buckles", locally relieve the strain in samples grown on $TiO_2$ $[001]_R$. This creates a gradual increase in $T_{IMT}$ on mesoscopic length scales as the center of the buckle is approached. The mid-infrared optical response of the highly inhomogeneous IMT in $VO_2$ films grown on $Al_2O_3$ has been previously characterized with static s-SNOM [17]. The character of emergent domains can be classified as being in the random field Ising universality class in a narrow temperature range surrounding the IMT [44]. Insights from ultrafast [8-10,45] and nanoscale [17] perspectives have also provided insight into the long-standing debate on whether the electronic or structural component is the driving force in the IMT, possibly revealing the existence of a monoclinic metallic state [46]. The first studies on nanoscale dynamics in $VO_2$ have recently been published [30,31].

Figures 1 b, and c summarize key experimental results. Here we plot data collected for $VO_2$ film on $TiO_2$ $[001]_R$ substrate. In Fig. 1b we plot the Tr-sHD signal, $\Delta S^{sHD} = (S_P^{sHD} - S_R^{sHD})/S_R^{sHD}$. In Fig 1c we plot the Tr-pHD signal, $\Delta S^{pHD} = (S_P^{pHD} - S_R^{pHD})/S_R^{pHD}$. With Tr-sHD a clear contrast is observed along buckles in our film (Fig. 1b) whereas this is not the case in data taken on a similar region of the $VO_2$ thin film using Tr-pHD under identical pumping conditions (Fig. 1c). We emphasize that no pump-induced features, which are above our noise floor, are observed in the data displayed in Fig. 1c. The discussion below will show that the result of Fig. 1c presents the genuine near field information whereas imaging in Fig. 1b is dominated by far-field artifacts. The results of Figs. 1b and c demand a critical evaluation of the data-acquisition protocols for time-resolved near-field measurements.

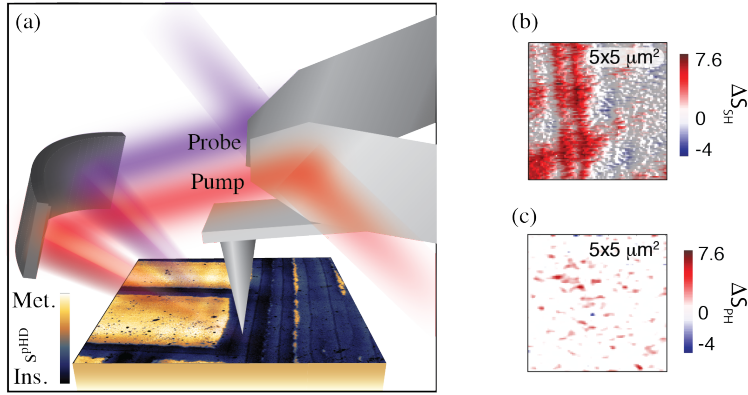

Fig. 1. Infrared Time-Resolved Nano-Imaging experiment and results. a) Diagram showing the Infrared Time-Resolved Nano-Imaging experimental apparatus. The ultrafast probe beam (purple) is focused onto the apex of an AFM probe at a precise time delay following a perturbation caused with a second ultrafast pump beam (red). Infrared nano-imaging data, which was collected with the Tr-pHD method using the 5$^{th}$ harmonic of the tip-tapping frequency. A pulsed laser source, whose center wavelength is 1.5 μm, was used to acquire the data shown on the $VO_2/TiO2$ [001] sample. This data, collected on a representative 10x10 μm$^2$ region, shows several metallic regions (gold) due to the compressive strain of the substrate as well as insulating regions (blue), where the film is strain relieved. b) Tr-sHD results obtained on the $VO_2/TiO2$ [001] sample in a 5x5 μm$^2$ region. c) Tr-pHD results obtained on the $VO_2/TiO2$ [001] sample in a 5x5 μm$^2$ region.

## 4. Methods for Time-Resolved Near-Field Detection.

We now proceed to develop a detailed comparison between the aforementioned data-acquisition protocols. Within the sHD method (Fig. 2a), one utilizes a 50/50 beam splitter to collect back-scattering radiation from the AFM probe and guide it to a detector. The AFM probe is tapped at a frequency $\Omega$ in the immediate proximity of the sample, which creates observable peaks at $n\Omega$ (with n = 1, 2, 3, etc.) when the detected signal is plotted against frequency (Fig. 2d). The HD method (Fig. 2b) requires the addition of a reference arm configured in an asymmetric Michelson interferometer scheme. The path length difference between the interferometer and the backscattered radiation from the AFM probe is set to zero such that both pulse trains interfere constructively, which enhances the signal at $n\Omega$ (Fig. 2e). The pHD method requires the path-length difference between the reference arm and sample to be modulated at a second frequency, M (Fig 2c). In pHD, the signal contained in the peaks at $n\Omega$ is partly transferred to sidebands separated by the tip tapping frequency, at $n\Omega +/- NM$, (where N=1, 2, etc.). As we will show in this section, the suppression of far-field background, and in turn the reliability of acquired data, is strongly affected by the choice of imaging method.

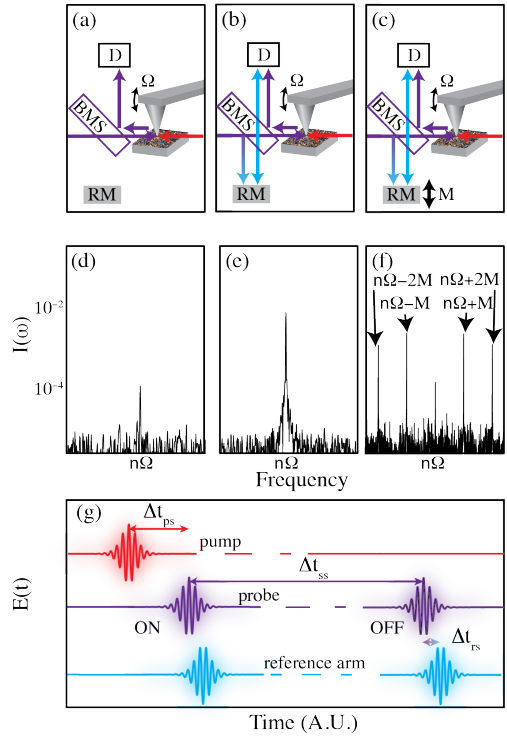

Fig. 2. Schematic of detection methods. a-c) Various detection methods with the radiation from the probe (purple), reference arm (blue) and pump (red) shown. BMS = 50/50 Beamsplitter; RM = Reference Mirror; D = Detector. a) sHD method, backscattered light from the AFM is steered into the detector. b) HD method where a reference arm is combined with the light from the self homodyne detection method. c) pHD method where the reference arm position is modulated at a frequency M. d-f) Signals acquired using the detection methods in a-c. d) sHD signal, which shows peaks at high harmonics of the tip tapping frequency $n\Omega$. e) HD signal, which shows that the magnitude of the peaks at $n\Omega$ are enhanced. f) pHD signal, which shows that, the peak at $n\Omega$ has returned to its sHD value and additionally peaks at the sum and or difference frequencies between the high harmonics of the tip tapping frequency and the reference arm $n\Omega +/- NM$ appear. g) Schematic of the pulses involved and relevant time scales. In the schematic we show the individual pump (red), probe (purple) and reference (blue) pulses on the femtosecond timescale. A much longer time delay, $\Delta t_{ss}$ – which is the inverse of the repetition rate of the laser system – is indicated by the dashed line. The dashed line separates the first (ON) event, where both the pump and probe pulses arrive at the sample and a second (OFF) event where only the probe pulse arrives at the sample. This process is periodically repeated, and data is collected by separately integrating the detected voltage from many ON and OFF events. In the case of HD and pHD methods radiation in the reference arm (blue) temporally overlaps with the probe radiation. In the case of the pHD method, the time delay between reference and probe light, $\Delta t_{rs}$ is modulated sinusoidally at a frequency M.

In Fig. 2g we show a schematic representation of the pulses involved and indicate the relevant time scales. To attain the highest possible signal-to-noise of the transient component of the near-field signal we adapted a boxcar based approach [41] to time-resolved s-SNOM measurements. In this approach one utilizes a pair of probe pulses. The first (ON) probe pulse follows the pumping event at time delay $\Delta t_{ps}$, marked by the red arrow in Fig.2g. This pulse provides signal associated with the pump-induced state of the sample at a time delay $\Delta t_{ps}$. An electro-optic, or acousto-optic, modulator is used to eliminate the second (OFF) pump pulse. The second (OFF) probe pulse, therefore, arrives at a much later time delay $\Delta t_{ps} + \Delta t_{ss}$ after the pumping event. Provided the sample has recovered its unperturbed state at $\Delta t_{ps} + \Delta t_{ss}$, the OFF signal contains information about the sample's unperturbed steady state. The intensity from both ON and OFF probe pulses are measured in a photoreceiver, whose response time is faster than the wait time between probe pulses, $\Delta t_{ss}$ (Appendix B), and the output is electronically integrated with a digital Boxcar (Zurich UHF-BOX). This process is repeated periodically and the integrated intensity values are registered as discrete data points at half of the repetition rate of the laser system. Standard lock-in demodulation of the boxcar output signals feeds the tapping harmonics of both ON and OFF probe pulses, provided that the repetition rate is sufficiently fast to satisfy the Nyquist criterion (Appendix B). The difference in the voltages demodulated from the ON and OFF pulses yields the information of reversible pump-induced changes to the sample.

Interferometric detection, where a pulsed laser is used, implies that the reference pulses temporally overlap with those from the sample on the detector. In Tr-HD, this is accomplished by using a micrometer stage to minimize the temporal mismatch between the tip-sample and reference arms ($\Delta t_{rs}$ in Fig. 2g), which places their interference at a constructive maximum. In Tr-pHD, the temporal-mismatch between probe and reference arms, $\Delta t_{rs}$, is first minimized and then modulated sinusoidally at a frequency M [34].

We now elaborate on the issue of multiplicative background fields, which was qualitatively raised in section 2, in a more rigorous fashion. This will allow us to extend our discussion to include time resolved measurements. Consider the intensity that is backscattered by an AFM probe, which is tapped at a frequency $\Omega$ (Fig. 1a). The electric field phasor that is backscattered from the tip and/or sample can be accurately described by a Fourier series expansion in harmonics of the tip-tapping frequency:

$$\tilde{E}_s = |\tilde{E}_0|e^{i\varphi_0} + |\tilde{E}_1|e^{i(\Omega t+\varphi_1)} + |\tilde{E}_2|e^{i(2\Omega t+\varphi_2)} + \cdots = \sum_n |\tilde{E}_n|e^{i(n\Omega t+\varphi_n)} \quad (3)$$

where $|\tilde{E}_n|$ (with n = 0, 1, 2, etc.) is the magnitude of the electric field at the n$^{th}$ harmonic of the tip-tapping frequency and $\varphi_n$ is the optical phase of scattered light encoded in the n$^{th}$ harmonic of the tip-tapping frequency. The leading term term, $\tilde{E}_0$, is largely unrelated to the tip-sample near field interaction so that this term may be accurately described as the background electric-field phasor, $\tilde{E}_{BG}$ (section 2). With increasing harmonic order the background contribution decays rapidly while the near-field contribution does not (Appendix A, Fig. 6). Thus, when high harmonics of the tip-tapping frequency, $\tilde{E}_n$, are accessed by demodulation of the detected intensity only terms that are proportionate to electric field phasor scattered by the near-field interaction, $\tilde{E}_{NF}$, contribute to the signal [34].

Apart from a select few detection methods, such as electro optic sampling and photo-conductive antenna detection, modern detectors measure the intensity of light rather than the electric field, as emphasized in Eq. (1). Since $\tilde{E}_{BG}$ dominates the high-harmonic component of signal by orders of magnitude the leading order term in the n$^{th}$ harmonic is [34]:

$$u_{sHD} \propto 2|\tilde{E}_{BG}||\tilde{E}_{NF}|\cos(\Delta\varphi_{BG}) \quad (4)$$

To simplify the equations in the remainder of the paper, we have introduced the phase difference $\Delta\varphi_{BG} = \varphi_{BG} - \varphi_{NF}$. In a pump-probe experiment one is exploring the difference between signals collected from the sample's pump induced states, at a time $\Delta t_{ps}$ following the pump pulse, and its static states. We, therefore, need to add an additional term to Eq. (4) to form the Tr-sHD signal:

$$\Delta u_{sHD}(x,\Delta t_{ps}) \propto |\tilde{E}_{BG}(\Delta t_{ps})||\tilde{E}_{NF}(x,\Delta t_{ps})|\cos\left(\Delta\varphi_{BG}(x,\Delta t_{ps})\right) -$$
$$|\tilde{E}_{BG}(\Delta t_{ps}<0)||\tilde{E}_{NF}(x,\Delta t_{ps}<0)|\cos\left(\Delta\varphi_{BG}(x,\Delta t_{ps}<0)\right) \quad (5)$$

It can immediately be appreciated that eight independent variables contribute to the Tr-sHD measurement; four of which are from the background electric field phasor. The presence of four BG variables, which all contain strictly diffraction limited information in Eq. (5) is discomforting. A cursory analysis of Eq. (5) reveals that the most troubling feature is an sHD experiment is not directly sensitive to the near-field phase. The coupled response between the, potentially, time dependent far-field phase and spatially dependent near-field phase can generate fictitious pump-induced features on the sub wavelength scale (Fig.1b). We note that this problem is not relevant if there are no pump-induced changes to the far-field phase. This complication is illustrated in in Fig. 3a and Table 1 for a specific choice of parameters that are relevant to the nanoscale dynamics of $VO_2$ in near-IR range (Appendix C).

A rather extreme, but plausible scenario, is displayed in Fig.3a: one single pixel in the field of view is characterized by a transient response that is different from other pixels in the same field of view. In Figure 3b we show the outcomes of this scenario evaluated with the help of Eq. (5). We see that a large ~4.5% Tr-sHD signal is predicted when the AFM probe is at the blue pixel. Furthermore, using reasonable experimental parameters for the optical constants of $VO_2$ (Appendix C), we have arrived at a Tr-sHD signal that is in approximate quantitative agreement with the measured near-infrared Tr-sHD results (Fig 1b).

By adding a reference arm it is possible to reduce, or completely suppress, the contribution of multiplicative background radiation (Appendix C). In Fig 3c we show the modeling results when a reference arm is added, which is known as homodyne detection Tr-HD (I) [39]. By considering typical intensity values for the reference arm relative to the sHD contribution we calculate that the magnitude of the fictitious change at the blue pixel should be reduced by a factor of 5x, as shown by the solid lines. By modulating the amplitude of the reference arm it is possible to fully suppress the contribution of the multiplicative background, which is known as heterodyne detection, Tr-HD (II) [36,37]. It is additionally possible to use a two-phase detection with Tr-HD (II) to extract the background-free near-field amplitude and phase [37], although this has not yet been applied to pulsed laser sources.

Likewise, when Tr-pHD is used, as in our current approach, one extracts near-field amplitudes and phases that are guaranteed to be artifact-free (sections 2, and 5).

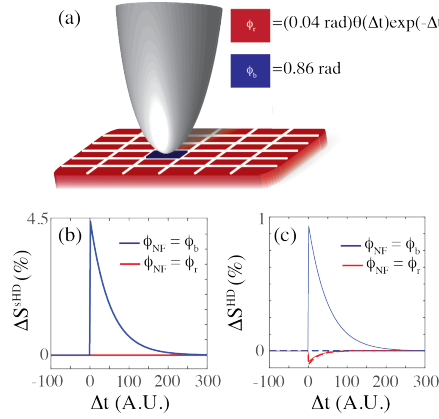

Fig. 3. Model calculations for the Tr-sHD and Tr-HD signals in near-IR. a) Schematic showing the AFM probe on a pixelated surface. The dominant area of the sample is state indicated with red, and is assigned the near-field phase $\phi_r$. A single pixel is blue, and is assigned the near-field phase $\phi_b$. The numeric values of these phases are shown in the inset. b) Transient response of the sHD signal at a red pixel (red) and blue pixel (blue). c) Transient response of the HD signal at a red pixel (red) and blue pixel (blue). The solid line shows the predictions for a typical ratio of reference arm to sHD intensities. The dashed line shows the case that the sHD intensity is set to zero, where the fictitious result at the blue pixel is removed.

|  | Actual | Tr-sHD | Tr-HD (I) | Tr-HD (II) | Tr-pHD |
|---|---|---|---|---|---|
| $\Delta s/s$ (red) | 0 | 0 | 0.1% | 0.15% | 0 |
| $\Delta\phi$ (red) | 0.04 rad | N/A | N/A | N/A | 0.04 rad |
| $\Delta s/s$ (blue) | 0 | 4.5% | 0.8% | 0 | 0 |
| $\Delta\phi$ (blue) | 0 | N/A | N/A | N/A | 0 |

Table 1. Model calculations for the Tr-sHD and Tr-HD signals for $VO_2$ film in near-IR range. The results of Tr-sHD were calculated using Eq. 5. The results of Tr-HD (I) were calculated using Eq. 6 with realistic magnitudes for the electric field of the reference arm relative to that of the background. The results of Tr-HD (II) were calculated using Eq. 6 with $|E_{BG}| = 0$. The values displayed for Tr-pHD can be obtained using Eq. 7-12.

We summarize the results of the above discussion in Table 1. The values shown are for the pump-induced changes in $VO_2$ film in near-IR range at the moment when the pump and probe arrive at the sample at the same time, $\Delta t_{ps}=0$. The genuine near-field pump-induced change in amplitude ($\Delta s/s$) and phase ($\Delta\phi$) are shown in the "actual" column for the red and blue pixel. It can readily be observed that only the near-field phase at the red pixel is non-zero. The calculated results of the Tr-sHD signal are shown in the third column. We see that in contrast to the "actual" change in near-field optical constants, the Tr-sHD signal is expected to show contrast at the blue pixel. When Tr-HD (I) is used false contrast is reduced by a factor of 5x and the time resolved response of the optical properties of the red pixel become observable. When Tr-HD (II) is used the fictitious signal at the blue pixel is fully suppressed while the finite signal at the red pixel is enhanced, which is in accordance with the "actual" scenario. Finally, when Tr-pHD is used, which will be described in the next section, measurements are expected to yield accurate results for the near-field amplitude and phase. We conclude that in the case that a detection method that is not affected by multiplicative background radiation such as Tr-pHD, Tr-HD (II), or EoS, are used results are guaranteed to be artifact-free.

## 5. Pseudoheterodyne detection for Artifact-Free Time-Resolved Near-Field imaging

In section 2 we have pointed out that the contribution from the background field is eliminated within the pHD method. To determine if the pHD method is compatible with pulsed laser sources, which are required for time resolved measurements, we proceed to discuss the quantitative details of the transient pseudoheterodyne method (Tr-pHD). In a static setting, pHD has been reliably used in a wide array of nano-infrared experiments over the past decade [17-19,34,47]. Within the pHD scheme the reference arm's phase is modulated at a frequency M. The reference arm modulation shifts the detection frequency from harmonics of the tip-tapping frequency, $n\Omega$, to sum and difference of tip-sample and reference arm frequencies, $n\Omega +/- NM$ (Fig 2a). As we explain in Appendix B, the demodulation involved in the extraction of a Tr-pHD signal is identical to the case that a C.W. laser is used. Therefore, the output of Tr-pHD is comprised of only terms where the reference arm and high harmonics of the AFM probe are mixed:

$$\Delta u_{pHD}(x, \Delta t) \propto |\tilde{E}_{ref}||\tilde{E}_{NF}(x, \Delta t)|\cos\left(\Delta\varphi_{ref}(x, \Delta t)\right) - |\tilde{E}_{ref}||\tilde{E}_{NF}(x, \Delta t < 0)|\cos\left(\Delta\varphi_{ref}(x, \Delta t < 0)\right) \quad (6)$$

Importantly, all dependencies of the Tr-pHD signal on the coordinate, x, or on the pump probe time delay, $\Delta t_{ps}$, enter Eq. (6) through the electric field that is scattered by the near-field interaction. For that reason, Tr-pHD produces genuine near field signal free of far field artifacts.

The second main benefit of Tr-pHD is that the amplitude and phase of the scattered field from the tip-sample interaction are simultaneously extracted, which has previously been shown for monochromatic laser sources [34]. The finite temporal duration of an ultrafast laser, however, implies that there is a finite bandwidth associated with the pulse train. In the case of broadband laser sources each frequency component of the detected is characterized with its own amplitude and phase. It is therefore prudent to examine the extraction of the pHD amplitudes and phases in the case of broadband laser sources.

In order to evaluate what signals are recorded by a lock-in amplifier, we consider the detected intensity, Eq. 1. The net intensity is comprised of three periodic events (1) the tip-tapping motion, (2) the reference arm phase modulation, (3) the arrival of laser pulses. We argue in Appendix B that the periodic train of laser pulses may be neglected in the demodulation of Tr-pHD signals. Therefore, we simply need to evaluate the appropriate Fourier expansion coefficients for the reference arm and tip-sample interaction [34]. We have given a mathematical expression for the electric field that is backscattered from the tip and/or sample in Eq. (3). The Fourier expansion coefficient of Eq. 3, is $c_n = |\tilde{\xi}_n(\omega)| e^{i\varphi_n(\omega)}$, where we emphasize the dependence of the electric field phasor on optical frequency ω [28]. To get the Fourier coefficient of the reference arm in the frequency domain we note that a sinusoidal variation in $\Delta t_{rs}$ (Fig. 2g) implies that the spectral phase is modulated as $\omega_0 a_m \cos(Mt)$ according to the Fourier shift theorem:

$$\tilde{\xi}_r(\omega) = |\tilde{\xi}_r(\omega)| e^{i(\omega a_m \cos(Mt) + \varphi_{ref}(\omega))} = \sum_{N=-\infty}^{\infty} |\tilde{\xi}_r(\omega)| J_N(\omega a_m) e^{i(\varphi_{ref}(\omega) + \frac{N\pi}{2})} e^{i(NMt)} \quad (7)$$

We used the Jacobi-Anger expansion in the second half of Eq. (7) to expand the reference arm electric field in terms of harmonics of the reference mirror modulation frequency. The Fourier expansion coefficient of the reference arm may be read directly from Eq. (7) as $c_N = |\tilde{\xi}_r(\omega)| J_N(\omega a_m) e^{i(\varphi_{ref}(\omega) + \frac{N\pi}{2})}$. Finally, as justified in Appendix B, when the detected voltage is demodulated at frequency $n\Omega$ +/ $NM$ the output is proportionate to the expansion coefficients, $u_{n,N} \propto c_N^* c_n + c_n^* c_N$.

In the case of a continuous wave (C.W.) laser, we evaluate $u_{n,N}$ at a single frequency, $\omega_0$:

$$u_{n,N} \propto |\tilde{\xi}_{NF}(\omega_0)||\tilde{\xi}_{ref}(\omega_0)| J_N(\omega_0 a_m) \cos\left(\Delta\varphi_{ref}(\omega_0) - \frac{N\pi}{2}\right) \quad (8)$$

Eq. (8) is identical to the formula for the detected voltage demodulated at frequency $n\Omega$ +/ $NM$ derived by Ocelic et al in Ref [34]. This equation can be further simplified when the first and second order Bessel functions are equal, i.e. when $J_1(\omega_0 a_m) = J_2(\omega_0 a_m)$, which happens with $\omega_0 a_m = 2.63$. This condition is satisfied by setting the reference mirror's physical amplitude to $\Delta l = c a_m / 4\pi\omega_0$. The amplitude of the near-field signal is then recovered by taking:

$$s_n \propto \sqrt{u_{n,1}^2 + u_{n,2}^2} \quad (9)$$

Likewise, the phase can be recovered with:

$$\varphi_n \propto atan_2(u_{n,2}/u_{n,1}) \quad (10)$$

where $atan_2(x)$ is the four quadrant inverse tangent of an argument x. This procedure provides a reliable method for extracting the near-field amplitude and phase in a static setting.

We now proceed to analyze the pHD method for signal recovery in measurements employing a broadband pulsed laser. An underlying assumption in data acquisition with a monochromatic source, shown above, is that the near-field amplitude and phase can be accurately expressed by a single numerical value. This assumption breaks down when a broadband source - where each frequency component has its own amplitude and phase - is used. In the case of a pulsed laser, which has a finite bandwidth, the condition that $J_1(\omega a_m) = J_2(\omega a_m)$ cannot be simultaneously satisfied for all $\omega$. We note that $a_m$ is a physical constant, which is set by the user of a near-field microscope. To determine the voltage recovered with a broadband source, one must perform an average over the frequency content of the laser, which is weighted by the frequency dependent response function of the detector element $\mathcal{R}(\omega)$ [28]:

$$u_{n,N} \propto \int \mathcal{R}(\omega)|\xi_{NF}(\omega)||\xi_{ref}(\omega)|J_N(\omega a_m) \cos\left(\Delta\varphi_{ref}(\omega) - \frac{N\pi}{2}\right) d\omega \quad (11).$$

A cursory inspection of Eq. (11) reveals that the finite bandwidth of the laser source introduces several complications. Near-field information collected is averaged over the bandwidth of the laser pulse, which is weighted by the frequency content in the magnitude of the intensity of the detected light and filtered by the spectral responsivity of the detector. This, however, is a consequence of the finite bandwidth of the source and cannot be avoided. The detected response is also weighted by the Bessel functions $J_N(\omega a_m)$ which is a specific consequence of the pHD method.

In the following text, we will analyze the uncertainty of both amplitude and phase measurements within the pHD method by considering the worst-case scenario error in experiments utilizing a broadband laser source. To do this, we calculate the pHD amplitude measured with a broadband source, Eq. (11), normalized to the monochromatic value:

$$s_n(\Delta\varphi_{ref})/s_{cw} = \frac{[(\int \xi(\omega)J_1(\omega a_m)\sin(\Delta\varphi_{ref})d\omega)^2 + (\int \xi(\omega)J_2(\omega a_m)\cos(\Delta\varphi_{ref})d\omega)^2]^{1/2}}{J_1(2.63\,rad)[\int \xi(\omega)d\omega)^2]^{1/2}} \quad (12)$$

In our calculations, shown in Fig. 4 we make several simplifying assumptions that yield a worst-case scenario estimate of the error. We consider a case that the spectral field, $\xi(\omega)$, is a box function in the frequency domain, shown in Fig. 4a, whose inverse Fourier transform has a pulsed nature (inset). We also neglect second order, and higher, spectral modifications to the phase so that $\Delta\varphi_{ref}$ is a constant. In Fig 4b we plot the normalized pHD amplitude Eq. (12), which is measured by a pulsed laser source, as a function of the laser bandwidth to center wavelength or relative bandwidth, $\Delta\omega/\omega_c$. We note that there are two shortcomings when pHD is used with a pulsed laser source. These are: (1) the amplitude that is recovered with a broadband laser source is less than what would be recovered with a monochromatic laser source; (2) the pHD amplitude recovered from a pulsed laser *does* depend on relative difference between the near-field phase of the sample and that from the reference arm, $\Delta\varphi_{ref}$. However, we observe that both of the aforementioned shortcomings are drastically reduced in the case that narrowband laser sources are used.

We proceed to quantitatively evaluate the extent to which the bandwidth of the pulse train may contaminate a measurement of the near-field amplitude. In Fig. 4b we display $s_n(0)/s_{cw}$ (black) and $s_n(\pi/2)/s_{cw}$ (red). We highlight in the inset that for a relative bandwidth, $\Delta\omega/\omega_c < 0.1$ the recovered amplitude is within a few percent of what the value that would be recovered for a monochromatic laser source, for all $\Delta\varphi_{ref}$, which is comparable to the smallest signals that are observable in state-of-the-art near-field experiments. For relative bandwidths less than 0.05 the error is less than 0.5% and will not be observable in most near-field measurements. Therefore, while there is a finite error in the amplitude recovery when pulsed laser sources are used with pHD, the error remains below typical noise levels in near-field experiments when narrowband laser sources are used.

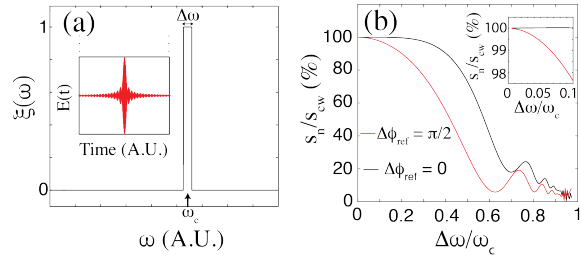

Fig. 4. Model of Experimental Error in Tr-pHD Amplitude. a) Spectral Field used in our calculation, $\xi(\omega)$, vs frequency, $\omega$. The Fourier transform of this Field, which has a pulsed nature, is displayed in the inset. b) Near-field amplitude collected with a pulsed laser normalized to the value that is anticipated for a monochromatic source, $s_n/s_{cw}$. We plot this quantity against the relative bandwidth of the laser source, $\Delta\omega/\omega_c$, as shown in panel a. We include this calculation for two values of the Near-field phase relative to that of the reference arm, $\Delta\phi_{ref}=0$ (black) and $\Delta\phi_{ref}=\pi/2$ (red). A zoom in of the narrow bandwidth region is shown in the inset.

The results presented in Fig. 4 show that the near-field amplitude and phase can be reliably recovered with pHD when a narrowband pulsed laser source is used. By incorporating a second illumination channel to pump the sample background-free time-resolved measurements of the near-field amplitude and phase may be carried out (red signal in Fig.'s 1a, 2g). When the boxcar approach is used to extract the Tr-pHD amplitude, Eq. (11), all common phase changes between the ON and OFF pulses are also canceled. These effects include drift of the reference arm phase relative to that from the sample, as well as static variations in the near-field phase (Section 4; Appendix B; Fig. 2g). Thus, the only changes observed in the Tr-pHD amplitude will stem from pump-induced changes to amplitude and phase of genuine time-resolved near-field features. This procedure provides a reliable method for extracting artifact free time-resolved near-field amplitude and phase.

## 6. Artifact-Free Time-Resolved Near-Field Results in Near-IR Range

In the previous sections we discussed advantages of the Tr-pHD method by modeling signals anticipated for $VO_2$ films in near-IR range. In Fig. 1 we have shown that while we detected a finite Tr-sHD signal in $VO_2$ films grown on $TiO_2$ [001], we were un-able to reproduce these results using Tr-pHD. We note that long timescales for recovery from the metallic state to the insulating state is characteristic of $VO_2$ films grown on substrates where the thermal conductivity is close to or less than that of the film itself [48]. These substrates include $SiO_2$ [31], and $TiO_2$ [19]. The recovery timescale can extend to hundreds of μs in films grown on these substrates, which supports the view [19] that cumulative heating may be the dominant pump-induced effect at our repetition rate of 300 kHz and may account for the observation of zero pump-induced near-field signal when the probe arrives hundreds of picoseconds after the pumping laser. To overcome this difficulty, we examined $VO_2$ films on substrates with thermal conductivity significantly higher than that of $VO_2$, where the recovery time is much less than 1.5 μs. These substrates include $Al_2O_3$, MgO, and Au.

In Fig. 5 we display the results obtained using Tr-pHD for a $VO_2$ film grown on a $Al_2O_3$ substrate. In Fig. 5a we plot the AFM topography. Grains are clearly observed in this image, which are typical of polycrystalline films [17,41,49]. In Fig. 5b we plot the static pHD data, where we observe slight variations in near-field signal at grain boundaries, which are probably due to a geometric modification to the local field enhancement. In Fig. 5c we plot the Tr-pHD signal, which is the normalized relative difference of the pHD signals taken from the pump-induced, $S_P^{pHD}$, and static, $S_R^{pHD}$, states, $\Delta S^{pHD} = (S_P^{pHD} - S_R^{pHD})/S_R^{pHD}$. We observe a ~2% homogeneous increase in the Tr-pHD signal at approximately pump-probe overlap, $\Delta t_{ps} = 0$. Our pump fluence is below the threshold required to fully excite the IMT. As reported in Ref. [30] when $VO_2$ is photo-excited with a fluence that is below the fluence threshold needed to generate an IMT, the pump-induced signal decays rapidly. The change in near-field signal decays by one order of magnitude within 1 ps [30,50]. These observations are consistent with our results.

We attribute the pump induced change in near-field signal at $\Delta t_{ps} = 0$ to the injection of free carriers into the conduction band. The pump-induced change to the near-field amplitude, which were collected with Tr-pHD, shows a completely homogeneous response at this time delay. The difference in our findings from Ref. [30] can stem from a number of factors including: the different wavelength of the probe, the differences between crystalline nano-beams utilized in Ref. [30] and granular films in Fig.5 or the different data acquisition method used.

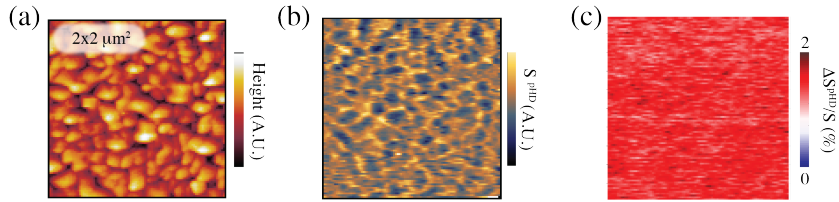

Fig. 5. Artifact-free Near-Field Data with a pulsed laser source. a) AFM data, which measures the topography, or local height, of the film in a 2x2 µm² region. b) pHD data with a pulsed laser source corresponding to the Topography in panel a. c) Tr-pHD data that was collected simultaneously with Figs 6 a and b.

## 7. Outlook and conclusions

In the header of Fig. 6 we briefly outline selected materials and phenomena that may be explored with pulsed laser sources in time-resolved and spectroscopic near-field measurements. At the longest wavelengths, THz s-SNOM is ideally suited to control and interrogate electronic properties [4,51], Josephson plasmon resonances in layered superconductors [52], hyperbolic polaritons in topological insulators [53], spin precession in ferromagnets [54] and anti-ferromagnets [55], as well as vibrational [56] and rotational [57] motions in a wide range of systems. The mid-IR spectral range is sensitive to the plasmonic modes in graphene [26,27,47,58,59], hyperbolicity in Hexagonal Boron Nitride [60], phonon resonances [35,41] as well as the electronic properties of many materials [17-19,41]. The pulse duration of mid-IR radiation is typically 40-200 fs, which is sufficient to gain access to timescales where electron-phonon, and electron-spin coupling have not yet brought the electronic system into thermal equilibrium [50]. In the visible range several interesting spectral features such as excitonic modes in transition metal dichalcogenides [61,62], plasmonic modes in metals and topological insulators [63], as well as resonances related to interband transitions in insulators, across charge transfer and Mott-Hubbard gaps can be observed. The pulse duration of visible radiation, which can be in the range of 4-40 fs, also enables indirect access to resonant modes in the infrared spectral range such as coherent phonons and Raman active modes [64,65]. Additionally, ultra short light pulses can be used for sub-cycle interrogation of processes excited with carrier envelope phase stable mid-IR [66] and THz pulses [55]. We stress that the efficient background suppression afforded by Tr-pHD with pulsed laser sources may find use in a wide array of spectroscopic measurements in addition to time-resolved control and interrogation of matter at the nanoscale.

In Fig. 6 we also display the calculated near-field signals and additive background contribution in these spectral ranges. These results show that as the demodulation order of the tip-tapping harmonic is increased the background contribution tends toward zero more quickly than the near-field signal over the entire spectral range plotted. We emphasize that that taking higher harmonics in the tapping technique does not eradicate multiplicative background artifacts (Section 2), and advanced techniques, such as Tr-pHD, must also be used to generate data that are completely immune from background radiation. The extremely high peak power densities, which are commonplace in pulsed laser sources, are ideal for exciting non-linear processes and enable generation radiation spanning from the EUV-THz with a single laser source. Thus, in addition to the time-resolved studies, which were at the focus of our current presentation, Tr-pHD with pulsed sources may find great utility in ultra-broadband, static, characterization of samples as well. Coupling pulsed sources to an s-SNOM enables novel opportunities for steady-state and time-resolved characterization of samples over an ultra-broad spectral range.

In the visible range achieving a excessive near-field to background ratio requires that very high harmonics are used, which in turn implies that there is a significant sacrifice to the achievable dynamic range in pristine nanoscale measurements. More exotic techniques that do not rely on the tapping technique have been demonstrated where all of the detected radiation stems from the near-field interaction [67,68]. These techniques have the capacity to preserve high S/N ratios, as well as ultra short pulse durations of broadband visible radiation, without compromising the high levels of background suppression that are required for proper near-field detection. The techniques in Refs [67,68] may, therefore, eventually provide a significant enhancement to the performance of Tr-SNOM experiments in the visible spectral range. In the mid-IR range the second or third harmonic provides nearly background free data. Interestingly, in the THz range even demodulation to linear order may provide adequate background suppression in many cases [51]. Under these conditions Tr-pHD can be used for reliable artifact-free time-resolved near-field imaging.

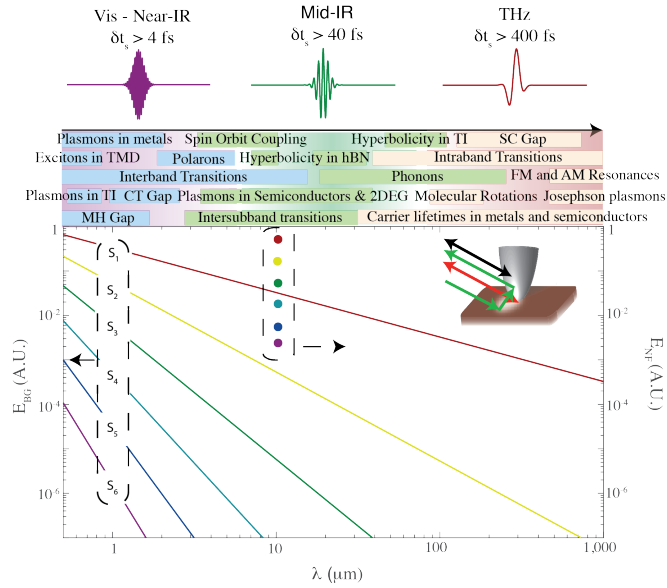

Fig. 6. Numerical values of the Near-Field and Background Contributions in s-SNOM measurements and the spectroscopic observables that may be explored with Tr-pHD. We identify three spectral regions. Vis - Near-IR where the temporal duration of laser pulses $\delta t_s$ is typically greater than 4 fs. Mid-IR where $\delta t_s$ is typically greater than 40 fs. THz where $\delta t_s$ is typically greater than 400 fs. Various spectroscopic observables are highlighted. TI = Topological Insulator; SC = Superconducting; TMD = Transition Metal Dichalcogenides; hBN = Hexagonal Boron nitride; FM = Ferromagnetic; AM = Anti-Ferromagnetic; CT = Charge Transfer; MH = Mott-Hubbard; 2DEG = 2D-Electron Gas. The plot shows the magnitude of the background electric field phasor (solid lines) calculated as described in Appendix A for harmonics of the tip-tapping frequency $s_1$ (red), $s_2$ (yellow), $s_3$ (green), $s_4$ (light blue), $s_5$ (dark blue), $s_6$ (purple). We also show the calculated magnitude of the electric field phasor from the near-field (dots) in the identical color scheme. In the inset we show a schematic representation of scattering processes that yield the background electric fields plotted here and discussed in Appendix A.

In conclusion, we have critically evaluated various detection protocols for time-resolved near-field measurements. Our modeling and experiments on VO$_2$ films show that the pHD method of acquiring transient pump-probe data that are guaranteed to eradiate complications arising from multiplicative background. It was shown that for narrowband pulsed laser sources ($\Delta\omega/\omega_c < 10\%$) pHD may be used in the same fashion as continuous wave laser sources - with the caveat that the pHD amplitude and phase recovered will be integrated over the bandwidth of the pulsed laser source. The limitations of, as well as novel time-resolved and spectroscopic possibilities using, Tr-pHD were detailed. Finally, we presented time-resolved nano-imaging data collected with the Tr-pHD method. The results indicate that Tr-pHD is a powerful tool for static and time-resolved nano-imaging and spectroscopy across a broad spectral range.

## 8. Funding and Acknowledgments

DNB is the Moore Foundation Investigator in Quantum Materials GBMF4533.

## Appendix A: Suppression of Background Radiation in the Tapping Technique

The nature of aperture-free near-field techniques provides an enhancement to the near-field contribution of the measured signal. As such it cannot be expected that the measured signal will be completely free from background radiation. It must, instead, be decided if the ratio between near-field and background information merits confidence that an arbitrary pHD signal is actually related to the material's response in the near-field. In the case of time resolved studies, experimental observables constitute a small fraction ($\Delta R/R = O(10^{-2}-10^{-6})$) of the overall signal. While the demand for adequate sensitivity in a pump probe experiment raises the bar for requirements on the signal-to-noise ratios of the near-field signal, the stringent requirement for adequate background suppression cannot be sacrificed to achieve a higher dynamic range. This section is intended to evaluate the possibility for time-resolved near-field signals to compete with time-resolved features from the background contribution.

In the inset of Fig. 6 we show a schematic layout intended to illustrate the origin of background contributions. The focal plane of the off-axis parabolic mirror, which is used to focus and collect light from the tip-sample interaction, can be brought above (black) or below (green) the plane of the sample (red). In each of these cases, radiation that interacts with the AFM probe can be back-scattered into the detector. As the AFM probe is lifted by a height $\Delta H$, the backscattered radiation experiences a phase shift of magnitude $\gamma = 2\pi \Delta H \cos\theta / \lambda$, where $\theta$ is the angle of incidence and $\lambda$ is the wavelength of probe radiation. Therefore, throughout the tip-tapping cycle, the phase of radiation that is backscattered directly from the probe is modulated as $\varphi_{bg} = \gamma \cos\Omega t$, with the tip-tapping frequency of $\Omega$. The background electric field is given by:

$$E_{BG} = |E|e^{i\gamma \cos\Omega t} = |E|\sum_{n=-\infty}^{\infty} J_n(\gamma)\, e^{i\frac{n\pi}{2}} e^{in\Omega t} \qquad (13)$$

Where we have again used the Jacobi-Anger expansion in the second half of Eq. (13). Eq. (13) shows immediately that background radiation will have a finite value at all harmonics of the tip tapping frequency. The background electric-field can, therefore, couple to the reference arm's electric field and produce a finite pHD signal in high harmonics of the tip-tapping frequency. While this source of background is an issue for static s-SNOM experiments, the background contribution in Eq. (13) is not affected by the pump-probe probe time delay, ($\Delta t_{ps}$ in fig 2g). This background source is, therefore, eliminated in Tr-pHD experiments.

It is also possible to measure a finite pHD signal from a background contribution that depends on the reflection coefficient of the sample. One situation in which this is possible is shown by the green beam in the inset of Fig. 6 where the focal plane of incident radiation is brought below the plane of the sample. In this case light that is reflected off of the sample, with reflection coefficient $r_{scatt}$, scatters off of the AFM probe shaft and is brought to the detector. By symmetry the phase shift experience by this reflected wave will be $\varphi_{bg} = -$

$\gamma cos\Omega t$ throughout the tip-tapping cycle. The background electric field in this case is given by:

$$E_{BG} = |E|r_{Scatt}(\Delta t_{ps})e^{-i\gamma cos\Omega t} = |E|\sum_{n=-\infty}^{\infty} r_{Scatt}(\Delta t_{ps})J_n(-\gamma)\, e^{i\frac{n\pi}{2}}e^{in\Omega t} \qquad (14)$$

In this case, it is possible to measure a finite Tr-pHD signal from the background contribution, since $r_{Scatt}$ is a function of the pump-probe time delay, $\Delta t_{ps}$. We note, however, that the background contribution is strictly diffraction limited and cannot vary on a deeply sub-wavelength length scale in real space.

In the case that the incident probe radiation is brought into the focal plane of the sample (red in the inset of Fig. 6), both aforementioned background sources contribute. Together with the bona fide scattered field associated with near-field interactions, the total electric field at high harmonics of the tip tapping frequency is given by:

$$E_n = E_{NF} + E_{BG} = |E_{NF}(x,\Delta t_{ps})|e^{i\varphi_{NF}(x,\Delta t_{ps})} + |E_{BG}|[J_n(\gamma) + r_{Scatt}(\Delta t_{ps})J_n(-\gamma)]e^{i\frac{n\pi}{2}} \quad (15)$$

Where we have explicitly noted dependencies on the local spatial coordinate, x, and the pump-probe time delay $\Delta t_{ps}$. The electric field in Eq. (15), which includes both near-field and background contributions is mixed with the reference arm to generate the pHD signal in a realistic experimental setting.

In static s-SNOM experiments, a sufficiently high harmonic is taken such that the background contribution becomes negligible with respect to the near-field signal. Starting from Eq. (15) we note that the tapping amplitude, ΔH, is chosen such that $\gamma = 2\pi\Delta H cos\theta/\lambda \ll 1$ to Taylor expand the Bessel functions to first order about zero. This gives us the total scattered electric field at the n$^{th}$ harmonic in a typical experimental setting [69]:

$$E_n \cong |E_{NF}(x,\Delta t_{ps})|e^{i\varphi_{NF}(x,\Delta t_{ps})} + |E_{BG}|\frac{(i\gamma)^n}{n!}(1 + (-1)^n r_{Scatt}(\Delta t_{ps})) \qquad (16)$$

It can be observed from Eq. (16) that the magnitude of the background term in a pHD signal will scale as $(\Delta H/\lambda)^n$. As the extent of the local evanescent wave that generates the near-field signal is approximately the tip-radius, ~20 nm, a tapping amplitude of $\Delta H \sim 60$ nm is sufficient to provide a large modulation of the near-field signal at an arbitrary wavelength. However, by keeping tapping amplitude constant, the degree to which the background radiation is affected by the tapping motion is strongly wavelength dependent. We show the background contribution for harmonics $S_1$ (Red) - $S_6$ (Purple) as a function of wavelength in Fig. 6. The magnitude of near-field signal as a series of increasing harmonic order was calculated at a wavelength of 10 μm using the lightning rod model [35] and is shown by the dots on the right hand side of the plot, with the identical color scale as the background contribution. We note that, the ratio of near-field signal to background signal should be nearly independent of wavelength, provided that the tip remains a good electrical conductor in the frequency range of the probe and that the wavelength remains much larger than the tip radius. The magnitude of the near-field signal relative to the background was normalized by using the experimentally measured ratio for the second harmonic at the probe wavelength of 1.5 μm, used in this work. This is the ratio of $s_2^{pHD}$ measured in when the AFM probe is in contact with the sample to the value of $s_2^{pHD}$ when the sample is fully retracted. The ratio of background to near-field contributions will, however, depend critically on the focused spot size as well as geometric factors so there may be significant error in the absolute comparison.

**Appendix B: Demodulation of the Pseudoheterodyne Signal using a Periodic pulse train of Femtosecond Light Pulses**

In the case of a periodic train of laser pulses, which are emitted at a repetition rate of $\Delta t_{ss}$ (Fig. 2c), the electric field phasor of a pulsed laser source is generally expressed as:

$$E(t) = \sum_{p\in\mathbb{Z}} |E(t - p\Delta t_{ss})|\exp\left[i\big(\omega t + \varphi(t - p\Delta t_{ss})\big)\right] \qquad (17)$$

Where the frequency of the laser source is given by $\omega$. The electric field magnitude and phase $|E(t - p\Delta t_{ss})|$, and $\varphi(t - p\Delta t_{ss})$ respectively, depend on the absolute time coordinate, t. The sum is over the set of all integers $\mathbb{Z}$. In a general case the electric field from the reference arm, and backscattered radiation from the AFM probe can have distinct time dependent amplitude and phase. Noting that the backscattered radiation is accurately expressed as a Fourier series in terms of the harmonics of the tapping frequency, the general form of the electric field phasor from the sample is:

$$E_n = \sum_{n\in\mathbb{Z}\geq 0} \sum_{p\in\mathbb{Z}} |E_n(t - p\Delta t_{ss})|\exp[i(\omega t + \varphi_n(t - p\Delta t_{ss}) + n\Omega t)] \quad (18)$$

In pHD, this is combined with a stronger electric field from the reference arm:

$$E_{ref} = \sum_{p\in\mathbb{Z}} |E_{ref}(t - p\Delta t_{ss})| \exp\left[i\left(\omega t + a_m \cos(NMt) + \varphi_{ref}(t - p\Delta t_{ss})\right)\right] = \sum_{N\in\mathbb{Z}} \sum_{p\in\mathbb{Z}} |E_{ref}(t - p\Delta t_{ss})|J_N(a_m)\exp\left[i\left(\omega t + NMt + N\pi/2 + \varphi_{ref}(t - p\Delta t_{ss})\right)\right] \quad (19)$$

The two electric field phasors combine to form an intensity, $I = E^*_{ref}E_n + E^*_{ref}E_n$, which is measured in a square law detector. Therefore, the full form of the intensity that is measured in pHD is:

$$I = \sum_{n\in\mathbb{Z}} \exp[-in\Omega t] \sum_{N\in\mathbb{Z}} J_N(a_m) \exp\left[\frac{iN\pi}{2}\right] \exp[iNMt] \sum_{p\in\mathbb{Z}} |E_k(t - p\Delta t_{ss})||E_{ref}(t - p\Delta t_{ss})| \exp\left[i\left(\varphi_{ref}(t - p\Delta t_{ss}) - \varphi_n(t - p\Delta t_{ss})\right)\right] + C.C. \quad (20)$$

Where we have introduced the repetition rate of the laser system, $f_{ss} = 1/\Delta t_{ss}$. To illustrate the influence of periodic laser pulse train we consider the simplified pulse structure - $E_n = \sum_{n\in\mathbb{Z}\geq 0} \sum_{p\in\mathbb{Z}} |E_n|\exp[\frac{(t-p\Delta t_{ss})^2}{T}]\exp[i(\omega t + \varphi_n + n\Omega t)]$ ; $E_{ref} = \sum_{p\in\mathbb{Z}} |E_{ref}|\exp[\frac{(t-p\Delta t_{ss})^2}{T}]\exp[i(\omega t + a_m \cos(NMt) + \varphi_{ref})]$. One can derive that Fourier coefficient of the pulse train is $c_p \propto \exp[A(\frac{p\Delta t_{ss}}{T})^2]$, where A is a constant of order unity.

Demodulation of the detected voltage at a frequency $pf_{ss} +/- n\Omega +/- NM$ will be proportional to $c_n c_N^* c_p + c_n c_N^* c_p^* + c_n^* c_N c_p + c_n^* c_N c_p^*$. Since we are considering pulses with $\Delta t_{ss} < 1\ ps$ and laser sources with repetition rates of $f$<1 GHz, $c_p$ is approximately constant for at least the first thousand harmonics of the laser repetition rate. We note that one must consider a convolution of this intensity with the temporal response of the detector, which evolves on a much slower timescale. Thus, we can safely assume that $c_p$ is a constant, which can be ignored in our notation that involves only proportionalities. One can observe that the detected voltage can be reduced to:

$$u_{n,N} \propto c_n c_N^* + c_n^* c_N \qquad (21)$$

For all frequencies, $n\Omega +/- NM$, which surround observable harmonics of the laser's repetition rate. It is then intuitively clear that in the case that a detector with a bandwidth $n\Omega < f_D < f$ is used, only the p=0, of this Fourier series survive. The extraction of the pHD signal is then identical to the case with a C.W. laser, so that we call this mode of operation quasi-C.W. If a detector with a bandwidth $f_D > f$ is used, the boxcar technique effectively performs a sum over all of the observable harmonics of the laser pulse train where pT<$f_D$. Each data point output by the boxcar integrator is, therefore, centered at zero frequency with a bandwidth $f_B$=f/2k, where k is the number of averages considered in the pulse train and the factor of 2 comes from the Nyquist criterion.

Therefore, when either quasi-CW operation, or boxcar based detection are used the Fourier expansion coefficients are identical to those published by Ocelic et al. [34], provided that the relevant harmonic for demodulation is contained within the bandwidth min($f_D$, $f_B$). This allows us to reduce the complexity of the Tr-pHD detection method and perform the common demodulation steps used for pHD with a CW laser source (i.e. demodulating at NM+/-n$\Omega$).

**Appendix C: Model of Artifacts in Tr-sHD and Tr-HD Measurements**

In this section we describe the technical details of the calculations used to quantitatively predict artifacts that are anticipated when Tr-sHD and Tr-HD methods are used, which are shown in Fig. 3a. In our model the majority of pixels, displayed in red, were assigned a static near-field phase $\varphi_r = 0$. We assign the phase of a single pixel, shown in blue, at $\varphi_b = 0.86\ rad$ relative to the static phase of light scattered from the red pixels, which is the measured phase difference between the insulating and metallic states of $VO_2$ at our probe wavelength of 1.5 μm. We also include a pump-induced change in near-field phase at the red pixels of $\varphi_r(\Delta t_{ps}) = (0.04\ rad)\theta(\Delta t_{ps}) \exp(-\frac{\Delta t_{ps}}{\tau})$, where $\theta(\Delta t_{ps})$ is the Heaviside function, $\tau$ is an arbitrary relaxation time constant, and $\Delta t_{ps}$ is the pump probe time delay. We emphasize that the magnitude of $(0.04\ rad)$ is modest for $VO_2$. The BG phase is simply the area averaged phase of near-field pixels, and thus $\varphi_{BG}(\Delta t_{ps}) \cong \varphi_r(\Delta t_{ps}) = (0.04\ rad)\theta(\Delta t_{ps}) \exp(-\frac{\Delta t_{ps}}{\tau})$. The anticipated results that are anticipated with various detection methods, summarized in Table 1, demonstrate that time-resolved signal that bears little to no relationship to the genuine optical constants provided are anticipated with Tr-sHD and Tr-HD methods. We therefore conclude, that the BG contribution to data gathered with Tr-sHD detection could produce the response shown in Fig 1b. and can generate observable artifacts in a general setting.

In Fig. 3c we consider the case that a reference arm is added (Figs. 2b, and e). In this case one measures:

$$\Delta u_{HD}(x, \Delta t_{ps}) \propto \left(|\tilde{E}_{BG}(\Delta t_{ps})|\cos(\Delta\varphi_{BG}) + |\tilde{E}_{ref}|\cos(\Delta\varphi_{ref})\right)|\tilde{E}_{NF}(x, \Delta t_{ps})| -$$
$$\left(|\tilde{E}_{BG}(\Delta t_{ps} < 0)|\cos(\Delta\varphi_{BG}) + |\tilde{E}_{ref}|\cos(\Delta\varphi_{ref})\right)|\tilde{E}_{NF}(x, \Delta t_{ps} < 0)| \quad (22)$$

Where we introduced, $\Delta\varphi_{ref} = \varphi_{ref} - \varphi_{NF}$. We emphasize that the radiation in the reference arm does not interact with the sample, thus the $|\tilde{E}_{ref}|$ and $|\varphi_{ref}|$ are properties that cannot depend on the temporal or spatial coordinates of the sample. Therefore, results generated with

Tr-HD signal will be valid if the amplitude of the reference field is much stronger than that of the background field, $|\tilde{E}_{ref}| \gg |\tilde{E}_{BG}|$ [39]. A quantitative estimate is also required to determine the possible contribution of background radiation in results generated with the Tr-HD method. In Fig. 3 and Table 1 we include such an estimate for near-IR range calculated using Eq. (22). In the case of Tr-HD (I) we use typical values for the magnitude electric field from the reference arm relative to that of the background as justified in Appendix C (solid lines). In the case of Tr-HD (II) we set the magnitude of the background electric field equal to zero, which can be accomplished in practice [36,37]. When Tr-HD (II) is combined with two-phase detection [38,47] the results are free from the influence of multiplicative background radiation and can be used to extract the amplitude and phase of the near-field signal, although this has not yet been demonstrated with pulsed laser sources. Other techniques, such as EoS or detection with a photo-conductive antenna are also immune from the multiplicative background contribution, and are compatible with rapid nano-imaging in the mid-infrared and terahertz frequency ranges [28,51]. Thus there is a whole arsenal of nano-optics methods that offer means and ways for multiplicative background radiation suppression.

By considering typical intensity values we are able to estimate the magnitude of measureable artifacts in Tr-HD. In a typical experiment the total light intensity backscattered by the AFM tip is approximately 4% of the incident intensity. The precise value may vary strongly, however, as this depends on many factors including the sample's roughness, probe beam-waist and collection efficiency of the off axis parabolic mirror. Radiation from reference arm comes from a mirror with 95-98% reflection. Therefore, without attenuation, the reference arm intensity outweighs the sHD intensity by approximately 25x. Since the electric field from the reference arm, rather than its intensity, enters into Eq. (22) a fully constructive interference between tip-sample and reference arms is approximately 5x greater than the sHD value. A fully constructive interference between HD detection and the near-field signal, in the majority of the sample, further implies that $\varphi_{ref} = \varphi_r$. We proceed to calculate the anticipated time-resolved response of the Homodyne signal using Eq. (22) with the parameters given in Fig 3a. In Fig 3c the solid lines show the result using an experimentally reasonable electric field in the reference arm. We observe that 1) the magnitude of the fictitious time-resolved response at the blue pixel is reduced by nearly one order of magnitude, and 2) a finite time-resolved response can now be observed at the red pixel. The artificial response at the blue pixel is, however, anticipated to remain dominant in a typical experimental setting. The dashed lines in Fig. 3c are obtained by setting $\Delta I_{sHD}(x, \Delta t) = 0$ in Eq. (22). The artificial time-resolved response at the blue pixel is eliminated only in the limit that the sHD signal is removed, while the time-resolved response at the red pixels remains observable. Thus, we conclude that artifacts in HD data are expected to remain significant in a typical experimental setting. In order to perform background free detection, it is essential to eliminate the contribution from Tr-sHD detection in the measured signal.